\documentclass[journal]{IEEEtran}
\usepackage{graphicx}
\usepackage{multirow}
\usepackage{amsmath}
\usepackage{algorithm}
\usepackage{algorithmic}
\usepackage{subfigure}
\usepackage[margin=2.5cm]{geometry}
\ifCLASSINFOpdf
\else
\fi
\begin{document}

\title{M-ATTEMPT: \\A New Energy-Efficient Routing Protocol in Wireless Body Area Sensor Networks}
\author{Nadeem Javaid~ \\ COMSATS Institute of Information Technology, Islamabad.}

\maketitle
\begin{abstract}
This paper presents an energy efficient routing algorithm for heterogeneous Wireless Body Area Sensor Networks (WBASNs). A prototype is defined for employing heterogeneous sensors on human body. Direct communication is used for real-time traffic (critical data) and on-demand data while multi-hop communication is used for normal data delivery in this proposed routing algorithm. One of the prime challenges in WBASNs is sensing of heat generated by implanted sensor nodes. The proposed routing algorithm is thermal-aware which sense the link Hot-spot and routes the data away from these links. Continuous mobility of human body causes disconnection between previous established links. We introduce mobility support and energy-management to overcome the problem of disconnection due to continuous mobility of human body. MATLAB simulations of proposed routing algorithm are performed for lifetime and reliability in comparison with multi-hop communication. The results show that the proposed routing algorithm has less energy consumption and more reliable as compared to multi-hop communication.
\end{abstract}


\section{Introduction}
\IEEEPARstart{P}{atient} monitoring is emerging as an important application of embedded sensors network. Many wireless sensors are implanted in/on the patient body. These tiny wireless sensors make Wireless Body Area Sensor Networks (WBASNs). A WBASN can observe physiological conditions of patient under supervision, and can provide us real-time feedback. Through WBASN a patient is constantly monitored, and in case of critical scenarios an immediate action should be required. These sensors can collect the physiological data, and this data is sent to physician in a hospital through Metropolitan Area Network (MAN) or Local Area Network (LAN). To diagnose the patient circumstances in real time and make a decision for it.

WBASNs are used for medical and nonmedical applications. The wireless sensor nodes used in WBASNs are tiny, light-weight and limited power sources. These sensor nodes have different levels of energy and generate different size of data while the Wireless sensor Networks (WSNs) nodes almost have same level of energy and data rate. Thus, employing routing algorithm of WSN can not support WBASNs sensor nodes. The selection of WBASNs routing algorithms should support the heterogeneous sensors network.

In \cite{1}, authors increase transmission range of sensor nodes and use single-hop communication between sensor nodes and sink node, to overcome the problem of topological partitioning due to constant human body movement and ultra short Radio Frequency (RF) transmission range.

Sue \textit{et al.} \cite{2} used multi-hop communication to transfer data between sink and root nods. In direct communication, increase in temperature of sensor nodes may affect human tissues. The storage delay (due to topological disconnections) and congestion delay increase delay in multi-hop communication. Delay is not supportive for emergency services.  Thus, Multi-hop communication cannot support emergency services.

In this paper, we present a mechanism for placing heterogenous sensor nodes on human body. We placed high data rate node on less mobile places on human body. Mobility of human body causes disconnection between previously established links. It takes time to establish new connection to forward data and causes delay. As delay is not supportive in real-time applications. To beat delay and overcome problem of disconnection. We used energy management in our proposed routing protocol. Using this energy management sensor nodes increase their transmission range and directly communicate with sink node for critical data delivery. For normal data delivery multi-hop communication is used.

In section II, we discuss a concise overview of background. We describe system model of proposed routing algorithm and also discuss different WBASNs employed sensor nodes and their data rates in section III. The simulation results and comparison of proposed routing algorithm with multi-hop algorithm is described in section IV. Finally the conclusion is drawn in section V.

\begin{figure*}[ht]
  \centering
 \subfigure{\includegraphics[height=08 cm, width=16 cm]{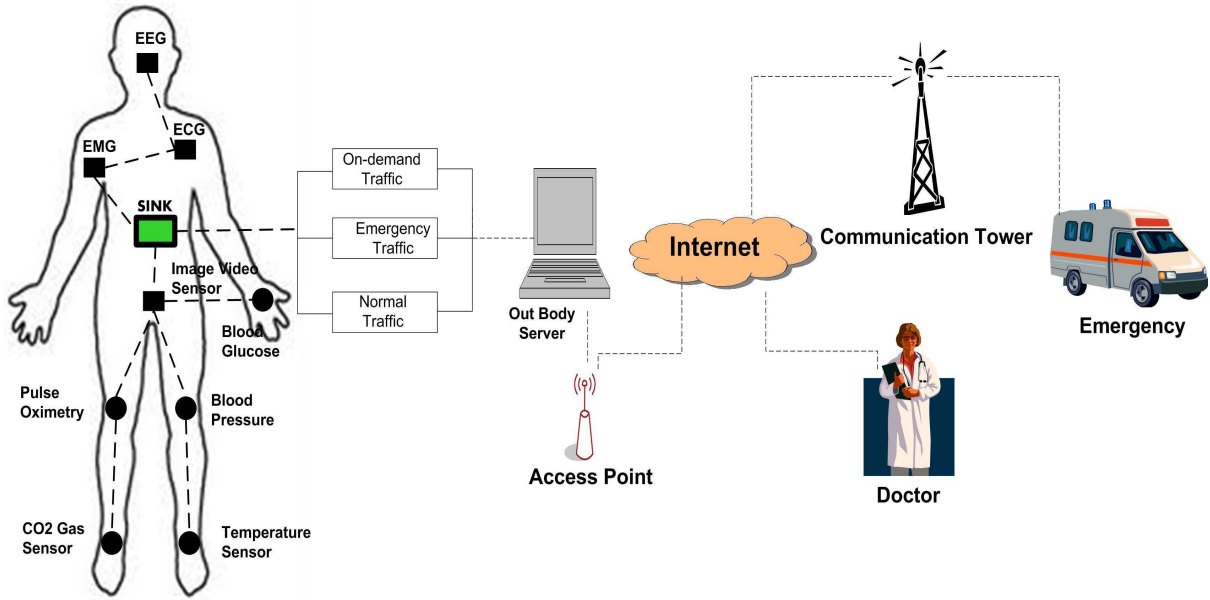}}
  \caption{Health-care application of WBASNs}\label{fig1}
\end{figure*}

\section{Background and motivation}
The rising temperature of implanted sensor nodes due to communication radiations and circuitry power consumption can affect the human body. In \cite{3}, authors use thermal-aware routing to minimize the effect of rising temperature of implanted sensor nodes. Quwaider \textit{et al.} \cite{4} use single-hop communication to transfer data between root nodes and sink node. They also define a prototype for placing sensor nodes on human body.

Environment Adaptive Routing (EAR) algorithm \cite{5} define different communication cost for heterogeneous WBASNs devices. However, the single-hop communication and proactive routing is not suitable for WBASNs. Multi-hop communication is suitable for normal packet delivery and single-hop is only used for emergency services due to high transmission cost. Use of Hello messages after a regular interval results high energy consumption.

In \cite{6}, Wireless Autonomous Spanning Tree Protocol (WASP)is defined to achieve low delay and network reliability for WBASNs. WASP-scheme message is disseminate to update parent nodes with information of child nodes. However, in WASP-scheme power balancing issue is not tackled.

Annur \textit{et al.} in \cite{7} apply tree algorithm with prioritization for WBASNs. A channel is dedicated for emergency data delivery and normal data is lagged until the successful delivery of critical data. However, the dedicated channel results loss of available resources.

In this work, we propose and validate a routing protocol both for heterogeneous and homogeneous networks. The nodes are placed around the sink in descending order of their data rate. As low data rate sensor nodes can not forward data generated by higher data rate sensor node due to insufficient buffer size. In the proposed protocol, priority is assigned to the emergency services. To overcome delay, critical data is sent directly to the sink node and normal data is sent through multi-hop communication. In next section, we discuss our proposed model in detail.

\section{System Model}
\begin{figure*}[ht]
  \centering
 \subfigure{\includegraphics[height=4 cm, width=16 cm]{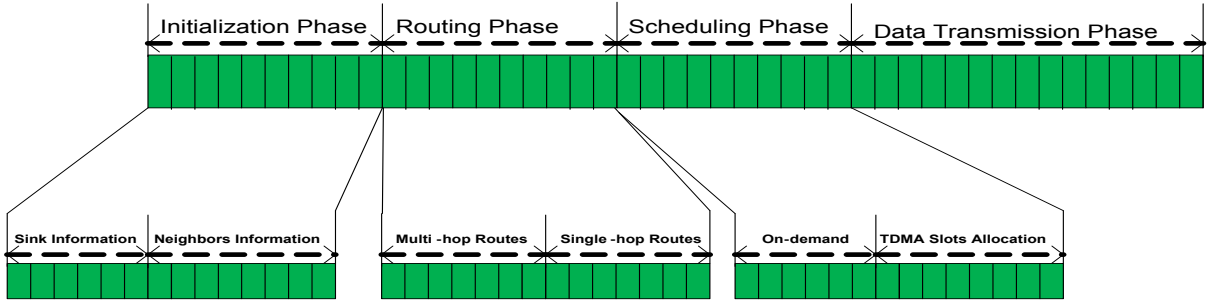}}
  \caption{Sequence of phases in each round}\label{fig2}
\end{figure*}

In order to introduce our model, we suppose that the sink is placed in the center of the human body. Since WBASNs are heterogeneous networks, and placement of nodes on human body is an issue. We resolved this issue by placing nodes in descending order of their data rate with respect to sink, as depicted in Fig. \ref{fig1}. Thus, the nodes with high data rate send data directly to the sink node, and can easily forward the received data from low data rate sensors. Implanted sensor nodes on human body with their data rate are mentioned Table. I. Problems analyzed in previous section are set in following manner: 1) single-hop communication is used for emergency services and on-demand data, 2) for normal data delivery multi-hop communication is used, 3) to prolong life-time of network by selecting the path with less hop-counts. Fig. \ref{fig2} depicts the phases of proposed routing protocol with above mentioned features. There are four phases in our in proposed routing protocol. These are initialization phase, routing phase, scheduling phase and steady state or data transmission phase. Initialization phase of the proposed routing algorithm is discussed in next subsection.

\begin{table}[!ht]
\begin{center}
  \begin{tabular}{| p{3cm} || p{3cm} |}
   \multicolumn{2}{c}{Table. I WBASNs sensors and their data rates}\\
  \hline
  \textbf{Sensors}                & \textbf{Data Rate}   \\ \hline \hline
   \textbf{EMG Sensor}            &  Very High	 \\ \hline
    \textbf{Image/ Video Sensor}  & Very High	 \\ \hline
     \textbf{Accelerometer}       & High \\ \hline
     \textbf{Blood Glucose}       & High	 \\ \hline
     \textbf{ECG Sensor}	      & High	 \\ \hline
      \textbf{EMG Sensor}         & High	 \\ \hline
      \textbf{Blood Pressure}     & Low	 \\ \hline
       \textbf{Tempraure Sensor}  & Very Low	 \\ \hline
      \textbf{$CO_{2}$ Gas Sensor}& Very Low	 \\ \hline
\end{tabular}
\end{center}
\end{table}

\subsection{Initialization Phase}
In initialization phase, all nodes broadcast Hello messages. These Hello message contains neighbors information and distance of sink nodes in form of hope-counts. In this way, all nodes are updated with their neighbors, sink node position and available routes to the sink node. Route computation for data delivery to sink node of the proposed routing algorithm is discussed in next subsection.

\subsection{Routing Phase}
In this phase, routes with fewer hopes to sink node are selected from available routes. We suppose nodes have information of all nodes and position of the sink nodes. So, selected routes are steadfast and consume less energy. Emergency services are also defined in proposed routing protocol. In critical scenarios, all processes are lagged untill successful reception of critical data at the sink node. In case of emergency, all the implanted nodes on the body can communicate directly to the base station. Moreover, all sensor nodes can communicate directly to the sink node when demand is arrived from sink node. In direct communication, delay is much less as compared to multi-hop communication. Because in multi-hop communication, each intermediate node receives, processes and then sends data to next node. The reception, processing and then transmitting the received data on each intermediate node which causes delay. And some time congestion also increased this delay. In critical scenario, delay is not acceptable. This delay is minimized by sending data through single-hop communication. We calculate energy consumed in single-hop communication $E_{S_HOP}$ as:

\begin{table*}[ht]
 \centering
  \begin{tabular}{| p{1.8cm} || p{1.8cm} || p{2.5cm} || p{1.8cm} || p{2cm} || p{1.8cm}|| p{1.8cm}|}
  \multicolumn{7}{c}{Table. II Comparison of ATTEMPT and M-ATTEMPT with existing algorithms}\\
  \hline
  \textbf{Algorithms}&\textbf{Network Type }&\textbf{Communication}&\textbf{Thermal-aware}&\textbf{Energy-efficient}&\textbf{Emergency}&\textbf{Mobility Support}  \\ \hline \hline
   \textbf{FPSS}[2] 	& Homogeneous   &   Multi-hop    &  Yes     & Yes   & Yes  & No    \\ \hline
   \textbf{TARA}[3]	    & Homogeneous   &   Multi-hop	 &  Yes     & No    & No   & No   \\ \hline
   \textbf{OBSFR}[4] 	& Homogeneous   &   Single-hop   &  No      & No    & No   & No   \\ \hline
   \textbf{EAR}[5]      & Heterogeneous & Multi-hop      &  No      & Yes   & No   & No   \\ \hline
   \textbf{WASP}\cite{6}	& Homogeneous   &    Multi-hop	 &  No      & No    & No   & No    \\ \hline
   \textbf{Tree} \cite{7}	& Homogeneous   &   Multi-hop	 &  No      & No    & Yes  & No     \\ \hline
   \textbf{DMQOS}\cite{razzaque2011data} 	& Homogeneous   &    Multi-hop	 &  No        & No  & No   & No      \\ \hline
   \textbf{ATTEMPT}	    & Heterogeneous & Single-hop/Multi-hop & Yes  & Yes & Yes  & No     \\ \hline
   \textbf{M-ATTEMPT}	    & Heterogeneous & Single-hop/Multi-hop & Yes  & Yes & Yes  & Yes     \\ \hline
\end{tabular}
\end{table*}

\begin{eqnarray}
  E_{S-HOP}=E_{transmit}
\end{eqnarray}
And transmission energy $ E_{transmit}$ is calculated as:
\begin{eqnarray}
  E_{transmit}= E_{elec}+ E_{amp}
\end{eqnarray}
where, $E_{elec}$ is the energy consumed for processing data and $E_{amp}$ is energy consumed by transmit amplifier. We suppose a linear network in which all nodes are implanted at equal distance from each other. To transmit $b$ bits up to $n$ hops, the transmission energy is given as:
\begin{eqnarray}
    E_{transmit}= n(b*E_{elec}+ b*E_{amp})* d^{2}
\end{eqnarray}
here $d^{2}$ is the energy loss due to the transmission.
\begin{eqnarray}
    E_{transmit}= n*b(E_{elec}+ E_{amp})* d^{2}
\end{eqnarray}
\begin{figure*}[ht]
  \centering
 \subfigure{\includegraphics[height=08 cm, width=16 cm]{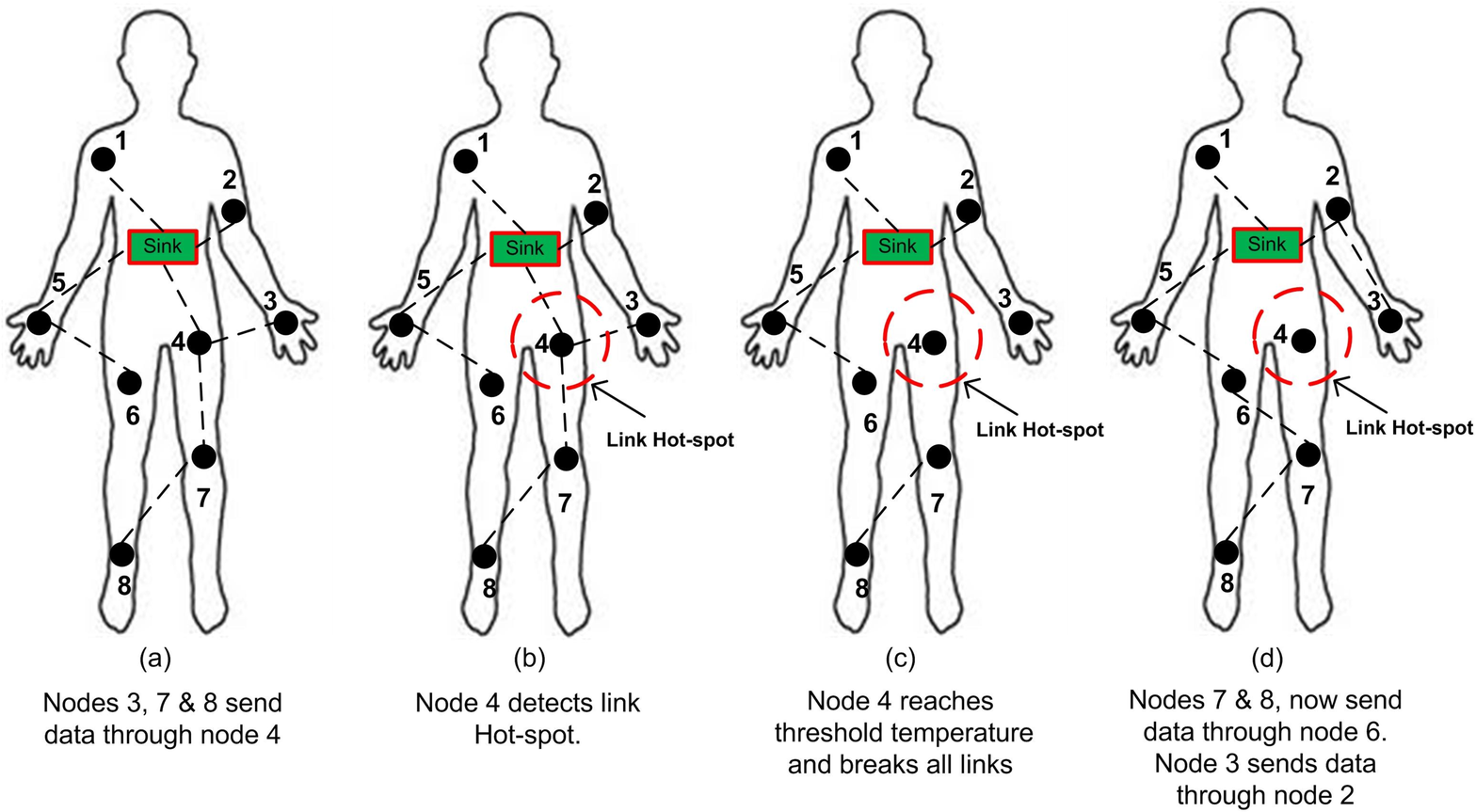}}
  \caption{Link Hot-spot detection}\label{fig3}
\end{figure*}

In our single-hop/multi-hop traffic control algorithm 1, if a node sensed emergency or on-demand data, then it uses single-hop communication. In single-hop communication, the sensor node uses full power of battery to sends its data. On the other hand, for normal data is received, the multi-hop communication is used to send data to sink node. Thus, collectively less power is consumed without effecting reliability in term of delay. As the distance increases more energy is consumed to send data. Thus, in multi-hop communication, energy consumption is very less.

Energy preservation is a prime consideration in WBASNs, as the deployed sensor nodes  have limited energy sources. So, deployed nodes need reasonable use of battery for extended life-time of the network. Implanted sensor nodes on human body have some heating effects.

\begin{algorithm}[H]
\caption{Single-hop/Multi-hop traffic control algorithm}
\begin{algorithmic}[1]
\FOR{i=1:1:n}
\STATE \textbf{Initialization Phase}
        \STATE T$_{r} \gets$ Transmission range of node i
        \IF{(Node$_{i}<=$ T$_{r}$)}
         \STATE Direct communication with Sink Node
        \ELSE
       \IF{(Critical Data ==1 $|$ On-demand==1)}
         \STATE Send data to Sink Node
                 \IF{( $L_{i,j}>$ C$_{T}$ )}
                   \STATE Send data to other route
                          \ENDIF
                             \ENDIF
                             \ENDIF
                             \ENDFOR
                          \end{algorithmic}
                           \end{algorithm}

 When we are dealing with wireless communication around the human body, effects of these sensor on human body can also be taken in to consideration. The most important factor consider for this purpose is Specific Absorption Rate (SAR) and heating effects of the implanted sensor nodes on human body. The purpose routing protocol is design to work according to SAR and heating effects on human body. As nodes implanted closer to sink node are forwarding data of their follower nodes. Whenever these nodes reach their temperature threshold, these nodes break their link to their neighbor nodes for few rounds. As their temperature become normal these sensor nodes establish their previous routes. However, if a sensor node receives a data packet and reaches its temperature threshold then it returns packet to previous node. And previous node mark this link as Hot-spot as shown in Fig.  \ref{fig3}. When we are dealing with normal data the Delay Tolerant Network (DTN) is supportive \cite{1}.

To calculate the energy consume during a multi-hop communication we assume a linear network in which all nodes are deployed at equal distance from each other. The loss of energy during multi-hop communication can be computed using the following equations:

\begin{eqnarray}
  E_{M-HOP}= E_{transmit}+E _{received}
\end{eqnarray}
\\
where, $E _{received}$ is the energy loss for receiving data. If we are transmitting $b$-bits to a distance of $n$-hops then the transmission energy will be $n*b*E_{transmit}$ and receiving energy will be $(n-1)b*E _{received}$. Since the first node transmit only and intermediate nodes first receive $n$-bits and then transmit these received bits. Therefore, the energy consumed for multi-hop is:

\begin{eqnarray}
E_{M-HOP} = n*b*E_{transmit}+ (n-1)b*E _{received}
\end{eqnarray}

from equation (2), equation (6) becomes:
\begin{eqnarray}
\begin{split}
E_{M-HOP} = n*b*(E_{elec}+ E_{amp}* d^{2})\\ + (n-1)b*E_{elec}
\end{split}
\end{eqnarray}

\begin{eqnarray}
\begin{split}
= n*b*E_{elec}+ n*b*E_{amp}*d^{2}\\+n*b*E_{elec}-b*E_{elec}
\end{split}
\end{eqnarray}

\begin{algorithm}[H]
\caption{: ATTEMPT Routing}
\begin{algorithmic}[1]
\STATE \textbf{Routing Phase}
            \IF{( route\_ 1 $<$ route\_  2 )}
                \STATE route\_  1 = selected route
              \ELSE
               \STATE route\_  2 = selected route
               \IF{( route 2 $<$ route\_  1 )}
                \STATE route\_  2 = selected route
              \ELSE
               \STATE route\_  1 = selected route
               \IF{( route \_ 1 = route \_ 2 )}
               \STATE $E_{hop-count} \gets$ Energy consumption for a route
               \IF{($E_{hop-count} \_1 < E_{hop-count}$ \_2 )}
                \STATE route\_  1 = selected route
              \ELSE
               \STATE route\_  2 = selected route
                          \ENDIF
                          \ENDIF
                          \ENDIF
                             \ENDIF
                          \end{algorithmic}
                           \end{algorithm}

\begin{eqnarray}
= [2*n*b*E_{elec}+ n*b*E_{amp}*d ^{2}-b*E_{elec}]
\end{eqnarray}

\begin{figure*}[ht]
  \centering
 \subfigure{\includegraphics[height=4 cm, width=16 cm]{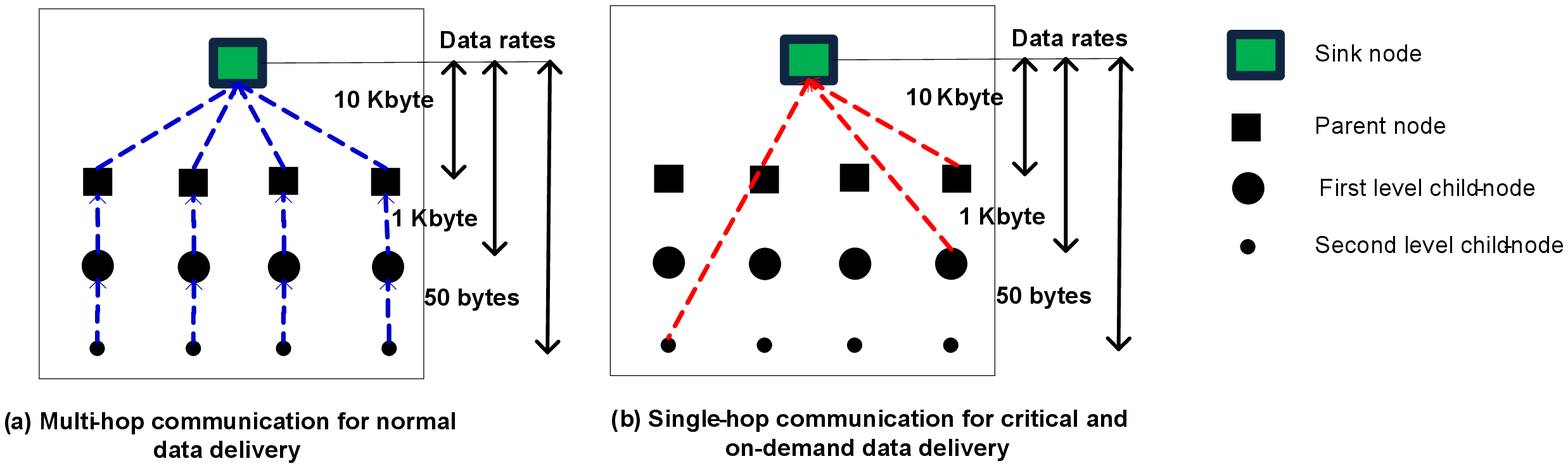}}
  \caption{Energy management for single-hop and multi-hop communication}\label{fig4}
\end{figure*}

Comparison of the attributes of proposed routing algorithm with existing routing algorithms are given in Table II. The ATTEMPT routing is discussed in Algorithm 2. If two routes are available the less hop-count route is selected. If two route have same hop-count, than route selected which have less energy consumption to the sink node. Single hop and multi-hop communication of root node with sink is shown in Fig. \ref{fig4}.

\subsection{Scheduling Phase}
After route selection phase, the sink node starts channel assignment using Time Division Multiple Access (TDMA) schedule for communication with root nodes. Sink node allocates time-slots to the root nodes for normal data delivery.

\subsection{Data Transmission Phase}
After time slots assignment to the root nodes, root nodes send their data to sink node in their allocated time slot. After that the sink node received data it will take some time to aggregate the received data. Sink node then send this aggregated data to the out body server through wireless link.	

\begin{figure}[ht]
\begin{center}
\includegraphics[scale=0.55]{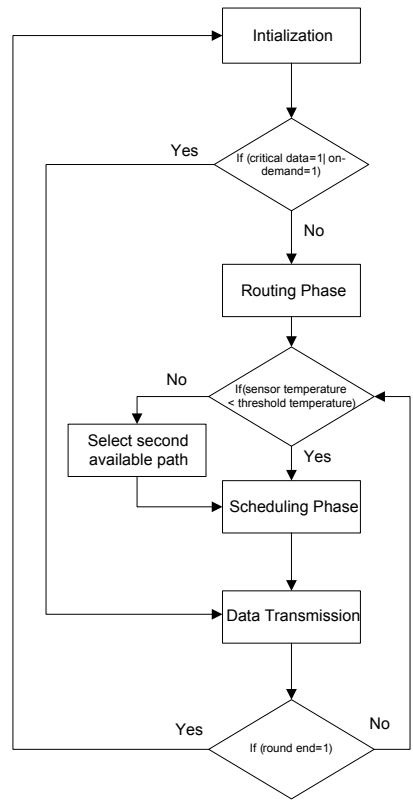}
\caption{Flow-chart of ATTEMPT}\label{fig5}
\end{center}
\end{figure}

\subsection{Mobility support in ATTEMPT}
WASNs are mobile in nature because of movements in human body. Our proposed scheme supports mobility. To achieve this we propose a prototype for placing nodes on human body and named it as Mobile-ATTEMPT (M-ATTEMPT). Nodes with high data rate are placed at less mobile places on human body. Theses nodes are parent nodes and directly connected to sink node. Parent nodes with $10J$ generate $10 Kbytes$ of data. The nodes directly connected to parent nodes are first level child-nodes with $5J$ generate $1 Kbytes$ of data. The nodes which are connected to first level nodes are second level child-nodes with $1J$ generates $50 bytes$ of data. Parent nodes, first level child-nodes and second level child nodes placed on human body and their respective topology. Overview of existing WBASNs routing protocols are discussed in \cite{ullah2010comprehensive} and are given in Table III. In next section, we discuss complete working of M-ATTEMPT. Sequence of phases during a round of M-ATTEMPT is depicted in Fig. \ref{fig7}.

\begin{figure*}[ht]
  \centering
 \subfigure{\includegraphics[height=12 cm, width=17 cm]{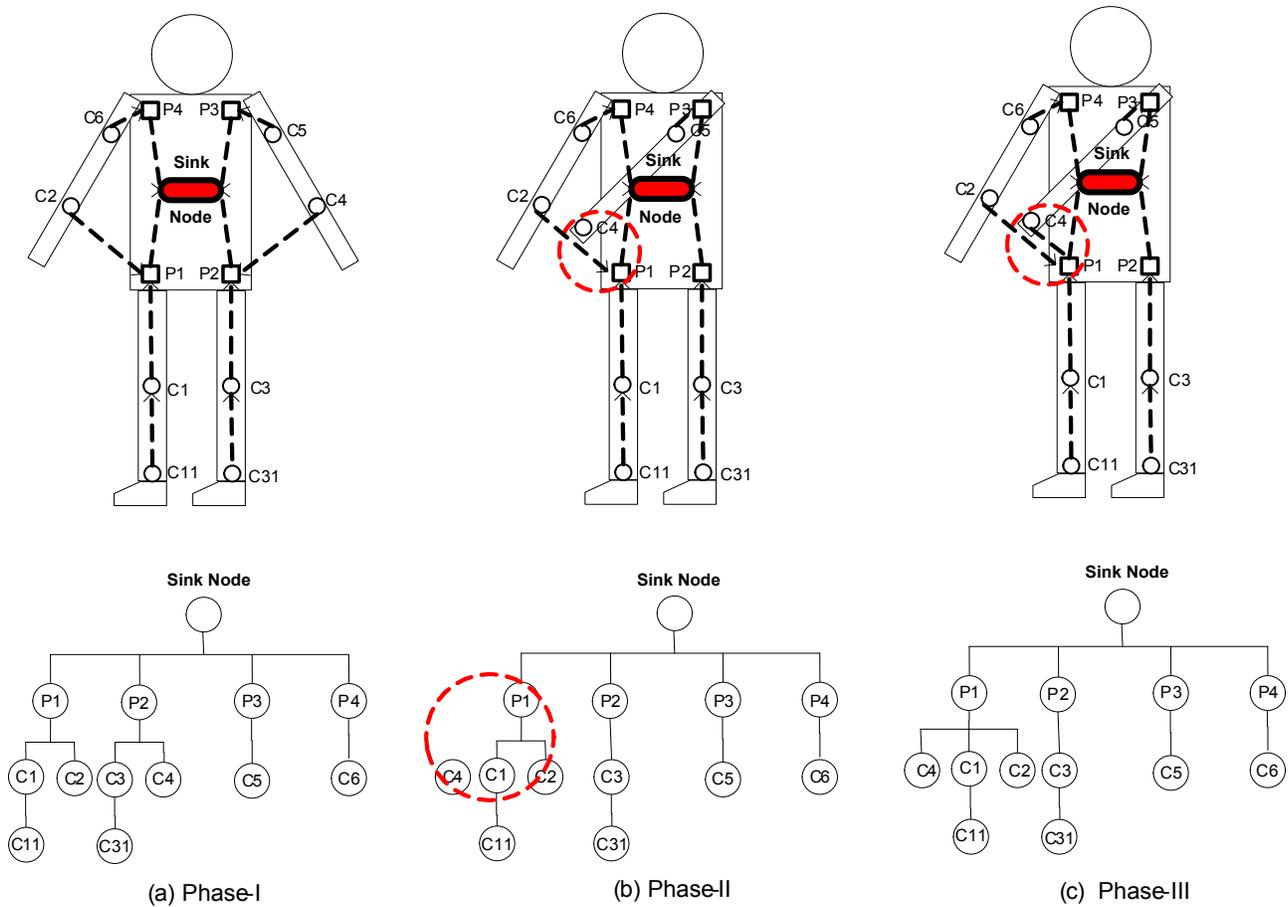}}
  \caption{Link establishment and link breakage due to mobility Of human body}\label{fig6}
\end{figure*}

\begin{figure*}[ht]
  \centering
 \subfigure{\includegraphics[height=5 cm, width=16 cm]{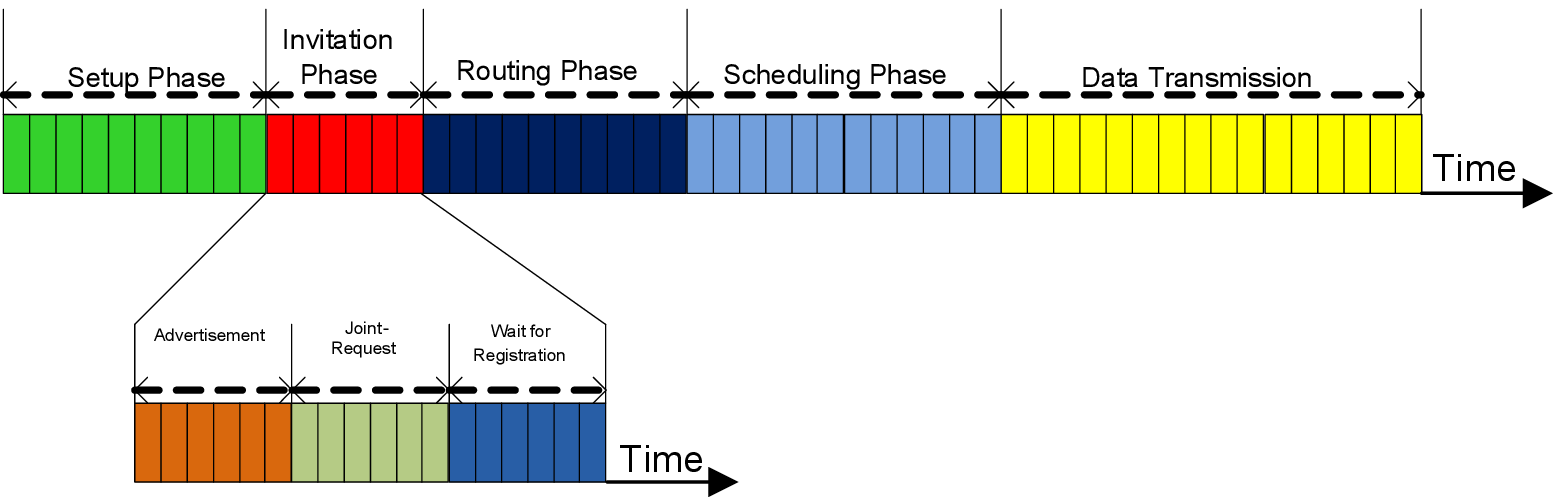}}
  \caption{Sequence of phases in each round}\label{fig7}
\end{figure*}

\begin{table}[!ht]
\begin{center}
  \begin{tabular}{| p{1.6cm} || p{1.5cm} || p{1.5cm} |}
   \multicolumn{2}{c}{Table. III Overview of WBASNs routing protocol}\\
  \hline
  \textbf{Cross Layer}                & \textbf{Temperature aware }&      \textbf{Cluster based}   \\ \hline \hline
   \textbf{WASP}                      & TARA	                   &       HIT \\ \hline
    \textbf{CICADA}                   & (A)LTR                     &       x\\ \hline
     \textbf{Ruzzelli}                & LTRT                       &       x\\ \hline

\end{tabular}
\end{center}
\end{table}


\subsection{Invitation Phase}
In this phase, we discuss how our proposed routing protocol support mobility. If a node changes its position during a round, nodes have to pay lot of energy to maintain link establishment. Sensor nodes only maintain its connection during mobility if sensor nodes sending an emergency data. As, human body changes its position, first level child node $C4$ disconnected from parent node $P2$ and entered in communication range of parent node $P1$, as depicted in Fig. \ref{fig6} phase-I. Now $C4$ will send joint-request to parent node $P1$, as shown Fig. \ref{fig6} phase-II. Parent node will check its parent child list, if number of child nodes are less then $\mu_{max}=3$. Then parent node $P1$ will accept joint-request and register $C4$ it as a child node as depicted in Fig. \ref{fig6} phase-III.
We have attenuation between root node and sink node which is proportional to $d^{2}$. The expression for this attenuation, $A$, as calculated in \cite{4726120}:
\begin{eqnarray}
  A_{j}=\sum^{n_{j}}_{i=1}d^{2}_{i}
\end{eqnarray}
Location of parent nodes can be computed from following equation:
\begin{eqnarray}
  X_{j}=\frac{1}{n_{j}}\sum^{n_{j}}_{i=1}X_{j}
\end{eqnarray}
\begin{eqnarray}
  Y_{j}=\frac{1}{n_{j}}\sum^{n_{j}}_{i=1}Y_{j}
\end{eqnarray}
Cost of energy paid by a root node during a round due to mobility of human body position is computed as:
\begin{eqnarray}
  C=v_{i}*\sqrt{(X_{i}-{X_{j})^{2}+(Y_{i}}-{Y_{j})^{2}}}
\end{eqnarray}
\begin{eqnarray}
v_{i}
\begin{cases}
v_{i}  \;\;\;\;\; if\;v_{i}<v_{t}\\
v_{i} \;\;\;\;\; otherwise
\end{cases}
\end{eqnarray}
\\
where, $v_{i}$ is velocity of mobile node, $X_{j}$ and $Y_{j}$ are defined in equation (19) and (20) respectively.
Some WBASNs sensor nodes with their data rate, bandwidth, power discussed in \cite{ullah2010comprehensive}, \cite{ latre2011survey} are given in Table IV.

\begin{table*}[ht]
 \centering
  \begin{tabular}{| p{3cm} || p{1.8cm} || p{2.5cm} || p{1.2cm} || p{1.5cm} || p{1.5cm}|| p{1.5cm}|}
  \multicolumn{7}{c}{Table. IV Comparison of different wearable sensors}\\
  \hline
  \textbf{Sensor nodes}&\textbf{Data type }&\textbf{Power consumption}&\textbf{QoS}&\textbf{Privacy}&\textbf{Accuracy}&\textbf{Band-width}  \\ \hline \hline
   \textbf{ECG} 	                & 288 kbps    &   Low    &  Yes      & High              & 12 bits  & 100-1000 Hz    \\ \hline
   \textbf{EMG}	                    & 300 kbps    &   Low	 &  Yes      & High              & 16 bits  & 0-10,000 Hz   \\ \hline
   \textbf{EEG} 	                & 43.2 kbps   &   Low    &  Yes      & High              & 12 bits  & 0-1 Hz         \\ \hline
   \textbf{Blood Pressure}          & 16 bps      &   High   &  Yes      & High              & 8 bits   & 0-150 Hz        \\ \hline
   \textbf{Temperature sensor } 	& 120 bps     &   Low	 &  Yes      & Extremely Low     & 8 bits   & 0-1 Hz          \\ \hline

\end{tabular}
\end{table*}
\section{Simulation Results}
We perform simulations to compare the performance of our proposed routing protocol with  multi-hop communication in MATLAB. We take network size of $5m$ x $5m$ in which $10$ nodes are randomly distributed and sink node is placed in the center of the network. We take $5000 rounds$ for these simulations. For mobility support we change positions of first level child-nodes and second level child nodes after $5 rounds$. All parameters taken for these simulations are given in Table V.

\begin{table}[!ht]
\begin{center}
  \begin{tabular}{| p{3cm} || p{3cm} |}
  \multicolumn{2}{c}{Table. V Simulation parameters}\\
  \hline
  \textbf{Parameters}         &  \textbf{Value}   \\ \hline \hline
   \textbf{Size of Network	} &  5 m x 5m	 \\ \hline
    \textbf{Number of Nodes}  &  10	         \\ \hline
     \textbf{Deployment}      &  Random      \\ \hline
      \textbf{Sink Location}  &  (2.5, 2.5)	 \\ \hline
   \textbf{Initial Energy}	  &  0.5 J	     \\ \hline
   \textbf{Number of Rounds}  &  5000	     \\ \hline
   \textbf{Application type}  &  Periodic-base/ Threshold-base	 \\ \hline
     \textbf{Packet Size}     &  $<=$ 64 Byte	 \\ \hline
     \textbf{Traffic Type}    &  CBR         \\ \hline
     \textbf{Radio Range}     &  $<=$ 10m	 \\ \hline
\end{tabular}
\end{center}
\end{table}

\begin{figure}[ht]
\begin{center}
\includegraphics[scale=0.45]{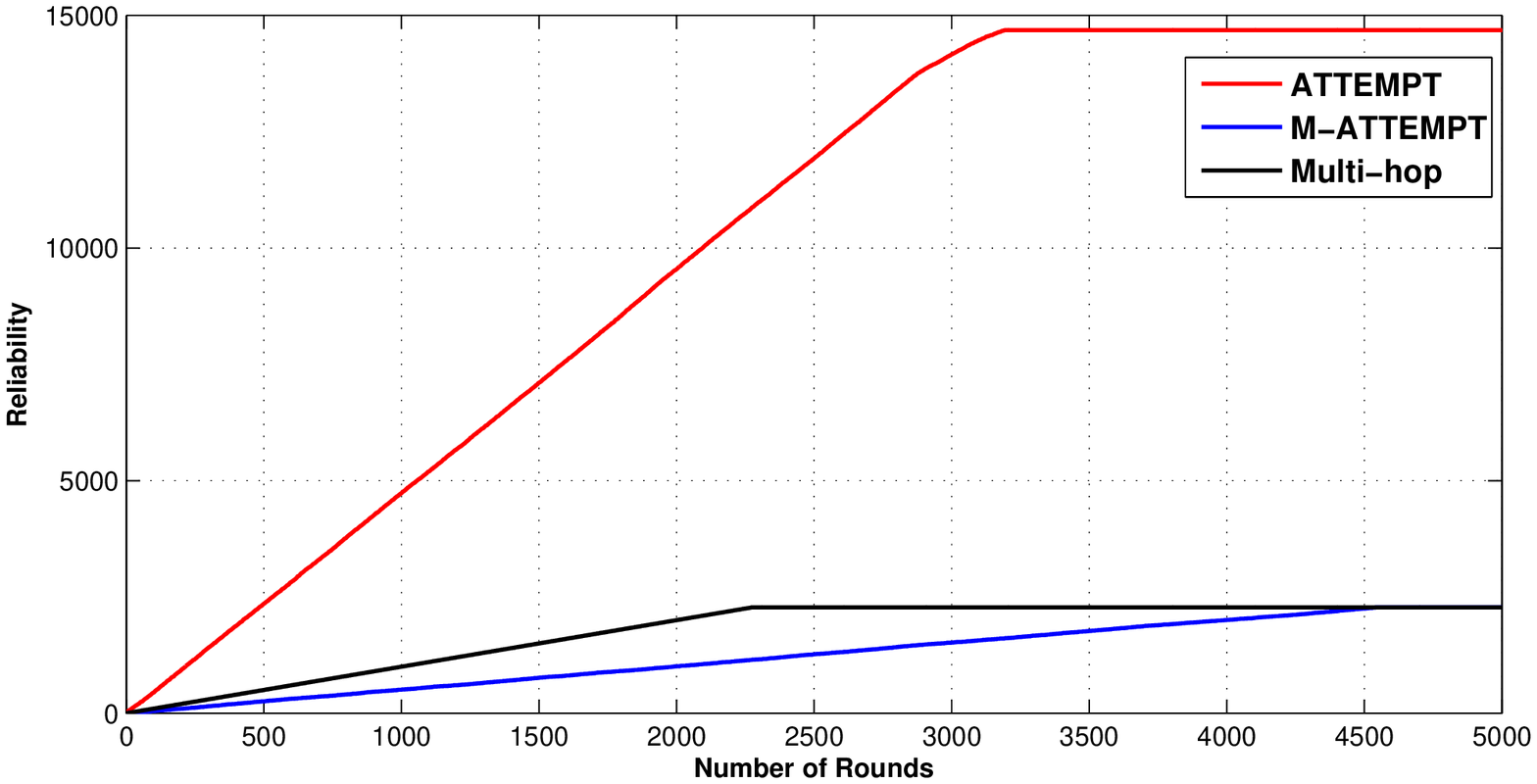}
\caption{Packet delivery ratio}\label{fig8}
\includegraphics[scale=0.45]{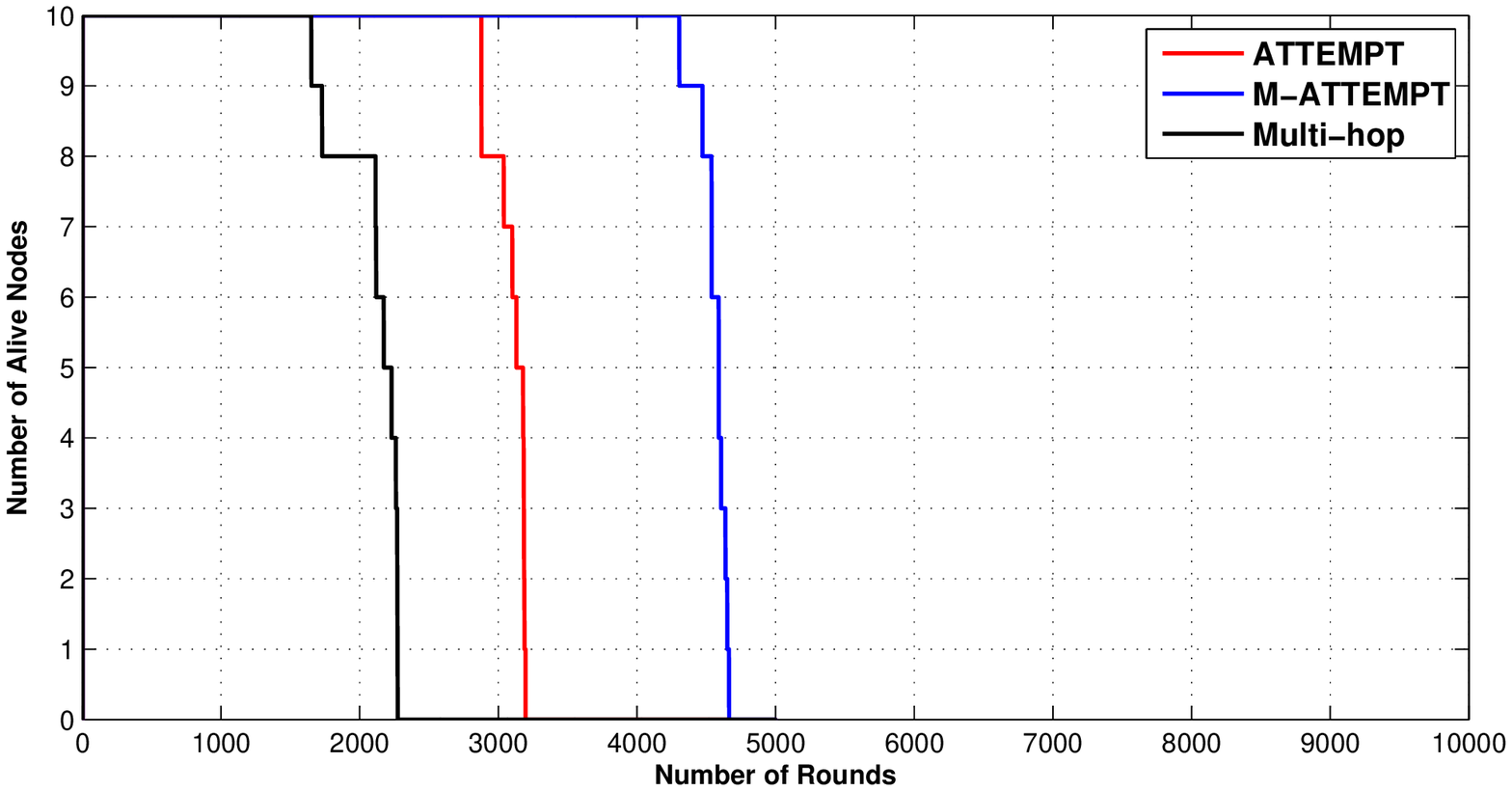}
\caption{Number of nodes alive over time }\label{fig9}
\includegraphics[scale=0.45]{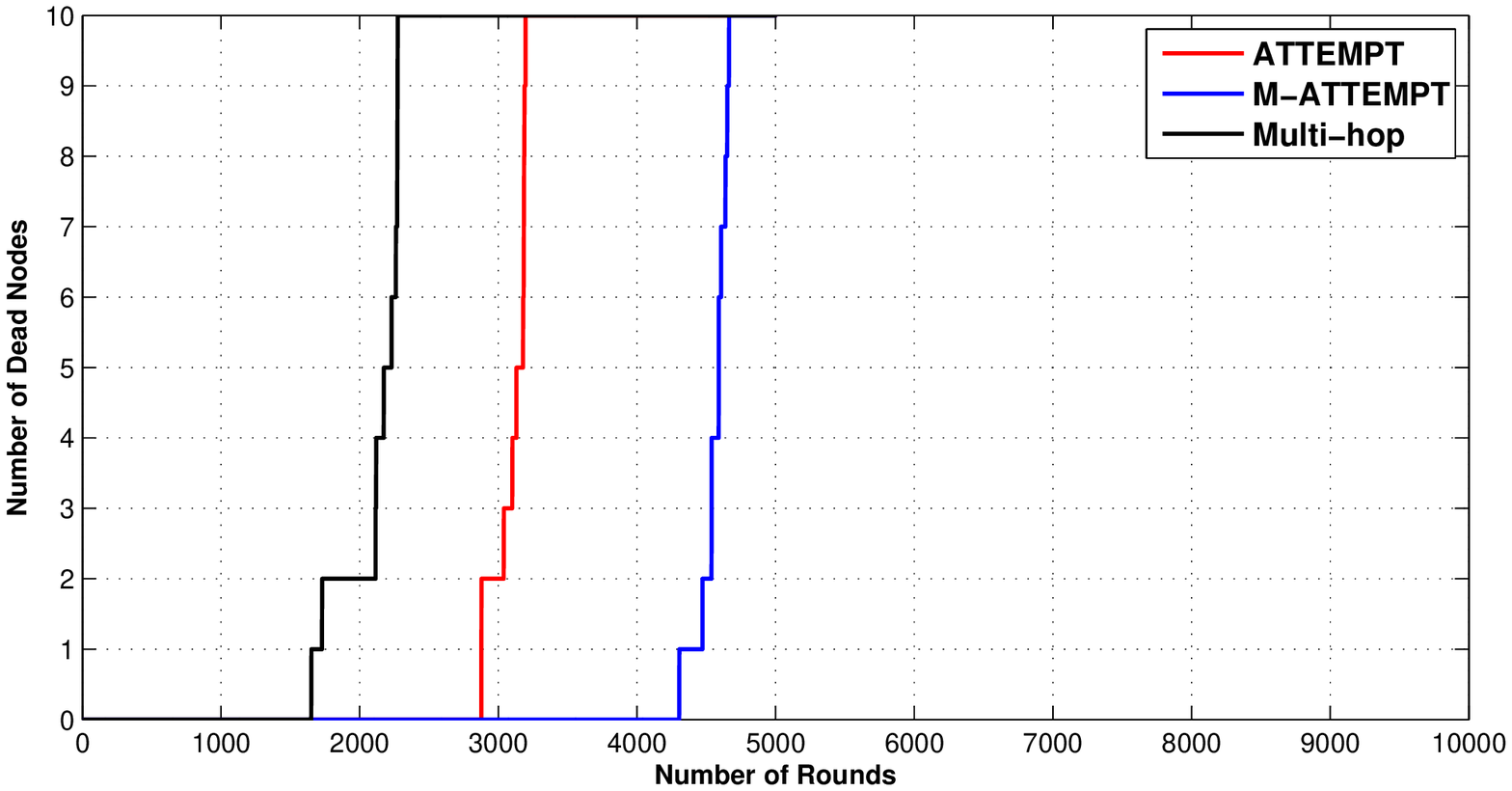}
\caption{Number of dead nodes over time }\label{fig10}
\includegraphics[scale=0.45]{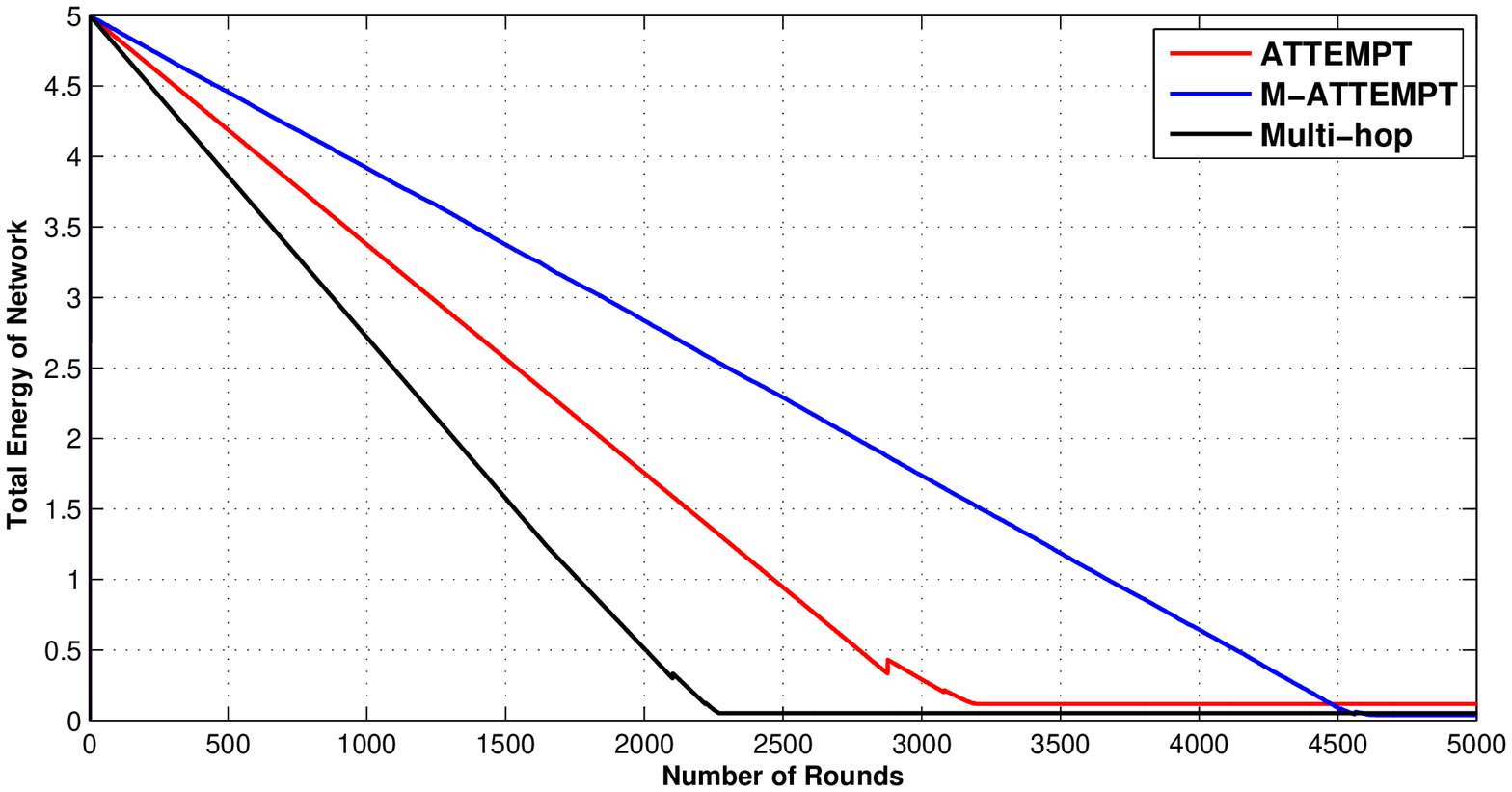}
\caption{Total energy of the network}\label{fig11}
\end{center}
\end{figure}

There is a complex trade-off between energy efficiency and fast routing in this mobile network. A multi-hop routing is the best choice for WBASNs and that one has to deal with a trade-off between energy efficiency and reliability. Reliability is experimentally investigated by measuring the packet delivery ratio. A multi-hop strategy turns out to be the most reliable. Fig. \ref{fig8} depicts that the reliability of ATTEMPT is almost $400\%$ better in stable and unstable region than M-ATTEMPT and multi-hop communication. The throughput of ATTEMPT is greater because it is sending threshold data and periodic data. While in case of M-ATTEMPT this packet drop also increased due to mobility of human body and less number of packets are receiving at sink node. However, as the mobility increased, packet drop rate increased. As a result of mobility fewer packets reached at sink node. As the distance and mobility of human body is increased between sink and deployed nodes the packet drop rate is also increased.

Stability period of a network is defined as when all nodes in a network are alive. Network lifetime of M-ATTEMPT is $48\%$ long as compared to multi-hop communication and almost $35\%$ greater lifetime, as compared to ATTEMPT. M-ATTEMPT has $32\%$ better stability period, as compared to Multi-hop routing and $20\%$ greater stability, as compared to ATTEMPT, as depicted in Fig. \ref{fig9}. Fig. \ref{fig10} shows instable period of network. M-ATTEMPT has $28\%$ less instable period, as compared to multi hop communication. The energy consumption of M-ATTEMPT is less and has better network lifetime, as compared to multi-hop communication and ATTEMPT. Total energy consumption of M-ATTEMPT, ATTEMPT and multi hop communication is depicted in  Fig. \ref{fig11}.

\section{Conclusion}
In this work, we present an energy efficient routing algorithm for heterogeneous WBASNs. For real-time and on-demand data traffic root node directly communicate with sink node and multi-hop communication is used for normal data delivery. Our proposed routing protocol supports mobility of human body with energy management. Sensor node increase and decrease their transmission range for single-hop and multi-hop communication to save energy. The proposed routing algorithm is thermal-aware which senses the link Hot-spot and routes the data away from these links. After selection of routes sink node creates TDMA schedule for communication between sink node and root nodes for normal data delivery. MATLAB simulations of proposed routing algorithm are performed for lifetime and reliability in comparison with multi-hop communication. Topology and placement of nodes is described with single-hop and multi-hop communication scenario. The results show that the proposed routing algorithm has less energy consumption and more reliable as compared to multi-hop communication.

\bibliographystyle{IEEEtran}

\bibliography{IEEEabrv,biblography}

\end{document}